\documentclass[journal=jacsat,manuscript=article]{achemso}

\usepackage[version=3]{mhchem} 
\usepackage{xcolor}
\usepackage{mciteplus}
\usepackage{chemfig}
\usepackage{caption}
\usepackage{amsmath}
\usepackage{graphicx}
\usepackage{ulem}
\usepackage{siunitx}

  \usepackage[english]{babel}
 \usepackage{float}

\author{Eladio Prieto Zamudio}
\affiliation[Michigan State University]
{Department of Chemistry, Michigan State University, 48824 East Lansing, MI, United States}

\author{Rituparna Das}
\affiliation[Michigan State University]
{Department of Chemistry, Michigan State University, 48824 East Lansing, MI, United States}

\author{Naga Krishnakanth Katturi}
\affiliation[Michigan State University]
{Department of Chemistry, Michigan State University, 48824 East Lansing, MI, United States}

\author{Jacob Stamm}
\affiliation[Michigan State University]
{Department of Chemistry, Michigan State University, 48824 East Lansing, MI, United States}

\author{Jesse Sandhu}
\affiliation[Michigan State University]
{Department of Chemistry, Michigan State University, 48824 East Lansing, MI, United States}

\author{Sung Kwon}
\affiliation[Michigan State University]
{Department of Chemistry, Michigan State University, 48824 East Lansing, MI, United States}

\author{Matthew Minasian}
\affiliation[Michigan State University]
{Department of Chemistry, Michigan State University, 48824 East Lansing, MI, United States}

\author{Marcos Dantus}
\affiliation[Michigan State University]
{Department of Chemistry, Michigan State University, 48824 East Lansing, MI, United States}
\alsoaffiliation[Michigan State University]
{Department of Physics and Astronomy, Michigan State University, 48824 East Lansing, MI, United States}
\alsoaffiliation[Michigan State University]
{Department of Electric and Computer Engineering, Michigan State University, 48824 East Lansing, MI, United States}
\email{dantus@chemistry.msu.edu}

\title{Enhanced strong-field ionization and fragmentation of methanol using non-commensurate fields}

\abbreviations{MS, SFI}
\keywords{femtochemistry, strong-field chemistry}

\begin{document}
\maketitle

\begin{abstract}
Electron-initiated chemistry with chemically relevant electron energies (10-200 eV) is at the heart of several high-energy processes and phenomena. To probe these dissociation and fragmentation reactions with femtosecond resolution requires the use of femtosecond lasers to induce ionization of the polyatomic molecules via electron rescattering. Here, we combine non-commensurate fields with intensity-difference spectra using methanol as a model system. Experimentally, we find orders of magnitude enhancement in several product ions of methanol when comparing coherent vs incoherent combinations of non-commensurate fields. This approach not only mitigates multiphoton ionization and multi-cycle effects during ionization but also enhances tunnel ionization and electron rescattering energy.
\end{abstract}

\section{Introduction}

Electron-initiated chemical reactions (EICR) are crucial for both fundamental and applied research, finding applications in diverse fields such as mass spectrometry, photolithography, combustion, plasma processing, atmospheric chemistry, ionizing-radiation medicine, and astrochemistry. Hence, there is a pressing need to move beyond a thermodynamic and statistical description to a molecular-level time-resolved understanding of EICR. Achieving femtosecond time resolution and energy resolution simultaneously has proven challenging for chemically relevant electron energies (10-200 eV), where the electron scattering cross-section is maximized. Fortunately, electron rescattering following tunnel ionization via irradiation with femtosecond near-IR pulses has emerged as a promising approach for achieving femtosecond and even attosecond time resolution for chemically relevant electron atom/molecule interactions.\cite{corkum1993plasma,niikura2002sub,li2020control} Given that the molecules themselves are the source of the electron that causes their own ionization, this approach allows for the time-resolved measurement of EICR.\cite{dantus2024tracking,dantussci2024} Unfortunately, the ionization process induced by a femtosecond laser pulse may involve unwanted effects (such as multiphoton ionization) depending on its energy, pulse duration, and wavelength.\cite{mics2005nonresonant,nairat2016order} These unwanted effects are less likely to induce electron rescattering, partially due to their relationship between ionization probability and laser phase. To address these challenges in EICR, we propose the interaction of a pair of femtosecond laser pulses with non-commensurate (NC) wavelengths to exploit the relatively phase-insensitive constructive interference in a minority of the optical cycles.\cite{bandulet2010gating,bruner2018robust,bruner2021control} Two wavelengths are called “non-commensurate” if their respective frequencies are not integer multiples of each other. Furthermore, these NC fields create sparse constructive inference optical cycles in the temporal intensity profile of the pulse. Additionally, we suggest employing intensity-difference spectra (IDS) to mitigate contributions from the leading and trailing wings of the laser pulses, as well as the lower energy regions of the Gaussian radial distribution of pulse energy, where multiphoton ionization is prevalent.\cite{posthumus2004dynamics,wang2005disentangling,wiese2019strong} Notably, for several pathways, we observe an order of magnitude discrimination when comparing ion yields obtained following excitation by the coherent compared to the incoherent sum of the two fields. This paper presents the theory and implementation of this approach, which we plan to utilize in future time-resolved studies.

The ionization of atoms and molecules using commensurate and NC fields has been an active field of research for decades. When combining the fundamental and second harmonic fields, the above-threshold ionization rate was enhanced and found to be phase sensitive.\cite{muller1990above} This inspired subsequent studies\cite{schafer1992phase,schumacher1994phase} to investigate the effects of relative phase-dependent forward-backward field asymmetry and the relative polarization \cite{fang2019strong,liu2018strong} on ionization yield,\cite{kotsina2015phase,kechaoglou2019controlling} the photoelectron momentum distribution,\cite{walt2015role,luo2017angular,madhusudhan2022strong} the directional ejection of strong-field ionized fragments,\cite{betsch2010directional,kaziannis2014interaction,song2015directional} and the control (enhancement) of the tunneling ionization and high-order harmonic generation.\cite{watanabe1994two, kondo1996tunneling, cormier2000optimizing,zhai2020ellipticity} The addition of NC wavelengths e.g. 1290 nm and 780 nm results in enhanced yield of mid-harmonics and higher cut-off energy as reported by Siegel et al.,\cite{siegel2010high} which was followed by Takahashi et al.\cite{takahashi2010infrared} and Lan et al.\cite{lan2010optimization} generating isolated attosecond pulses with a continuum high-order harmonic spectra\cite{bandulet2010gating} using multi-cycle pulses with carrier-envelope phase (CEP) insensitivity by optimizing the wavelength of the assisting field. Due to the constructive and destructive interference between the NC fields, the resultant field consists of well-separated optical cycles with altered tunneling ionization rates and subsequent sub-cycle electron dynamics.\cite{bruner2018robust,bruner2021control} While NC fields have been shown to be useful by themselves, ideal energy resolution requires the elimination of unwanted field effects from the femtosecond pulses. This is why we combine the advantages of using NC fields with the IDS method,\cite{posthumus2004dynamics,wang2005disentangling,wiese2019strong} which accounts for the focal volume effect arising from the molecular ensemble interacting with a laser pulse which is Gaussian both spatially and temporally. This combination enables us to distinguish between fragmentation pathways arising from tunnel ionization and multiphoton ionization.

We focus on the dissociative ionization of methanol as a model polyatomic system, which has been extensively studied by electron impact with energies ranging from 10 to 500 eV.\cite{Charles1940,Szmytkowski1995,Satyendra2004,Silva2010,Varela2015,Nixon2016} This molecule is used as a model for molecules involving \ce{O}, \ce{C}, and \ce{H} in the interstellar medium\cite{Vinodkumar2008} and it has been fundamental in studying the formation and reaction of \ce{H3+}.\cite{Burrows1979, Sankar2006,Dantus2017, Dantus2018, Strasser2020}  Here, we apply femtosecond NC fields and IDS to the strong-field ionization of methanol, in order to study the fragmentation in isolated regions of single electron tunnel rescattering without contributions from other mechanisms, such as multiphoton ionization.

\section{Theoretical Concept}

We now investigate the dependence of the strong-field ionization process on the temporal characteristics of the NC fields. Mathematically, the NC field can be represented as:

\begin{equation}
        E(t) = \sqrt{\frac{2}{\epsilon_0c}}e^{\frac{-t^2}{\tau^2}} \{ E_1cos(\omega_1t+\phi_1) +  E_2cos(\omega_2(t - \Delta t)+\phi_2)\}
\end{equation}

where $E(t)$ is the resultant NC field, $c$ and $\epsilon_0$ are fundamental constants. $\tau$ is a constant related to the pulse duration, $E_1$ and $E_2$ are the electric field amplitudes of each pulse, and the $\omega$'s and $\phi$'s represent the central angular frequency and carrier-envelope phase for each pulse, respectively. $\Delta t$ is the time delay between pulses. With this description of the NC fields, we can begin predicting the regime of ionization that is expected to be dominant at different points of the field. This is commonly applied to atoms using the Keldysh parameter:

\begin{equation}
    \gamma = \sqrt{\frac{I_p}{2U_p}}
\end{equation}

This parameter helps delineate between the two limiting cases of strong field ionization: the multiphoton regime ($\gamma >> 1$) and the tunneling ionization regime ($\gamma << 1$). It is this second regime that we wish to isolate to study electron initiated chemistry. Once this regime is isolated, one can compute the energy incident on the molecule via the rescattered electron using the equation for the ponderomotive energy:

\begin{equation}
\label{Eq2}
    U_p=\frac{e^2 E^2}{4m\omega_0^2}
\end{equation}

where the rescattered electron can achieve at maximum a constant multiple of this energy (3.17 $U_p$) for a monochromatic field.\cite{corkum1993plasma} Note that to achieve the maximal electron kinetic energies, one wishes to decrease the frequency of the ionizing field or combine multiple frequencies as is done here to raise this kinetic energy ceiling.

 \begin{figure}[H]
 \centering
   \includegraphics[width=\textwidth]{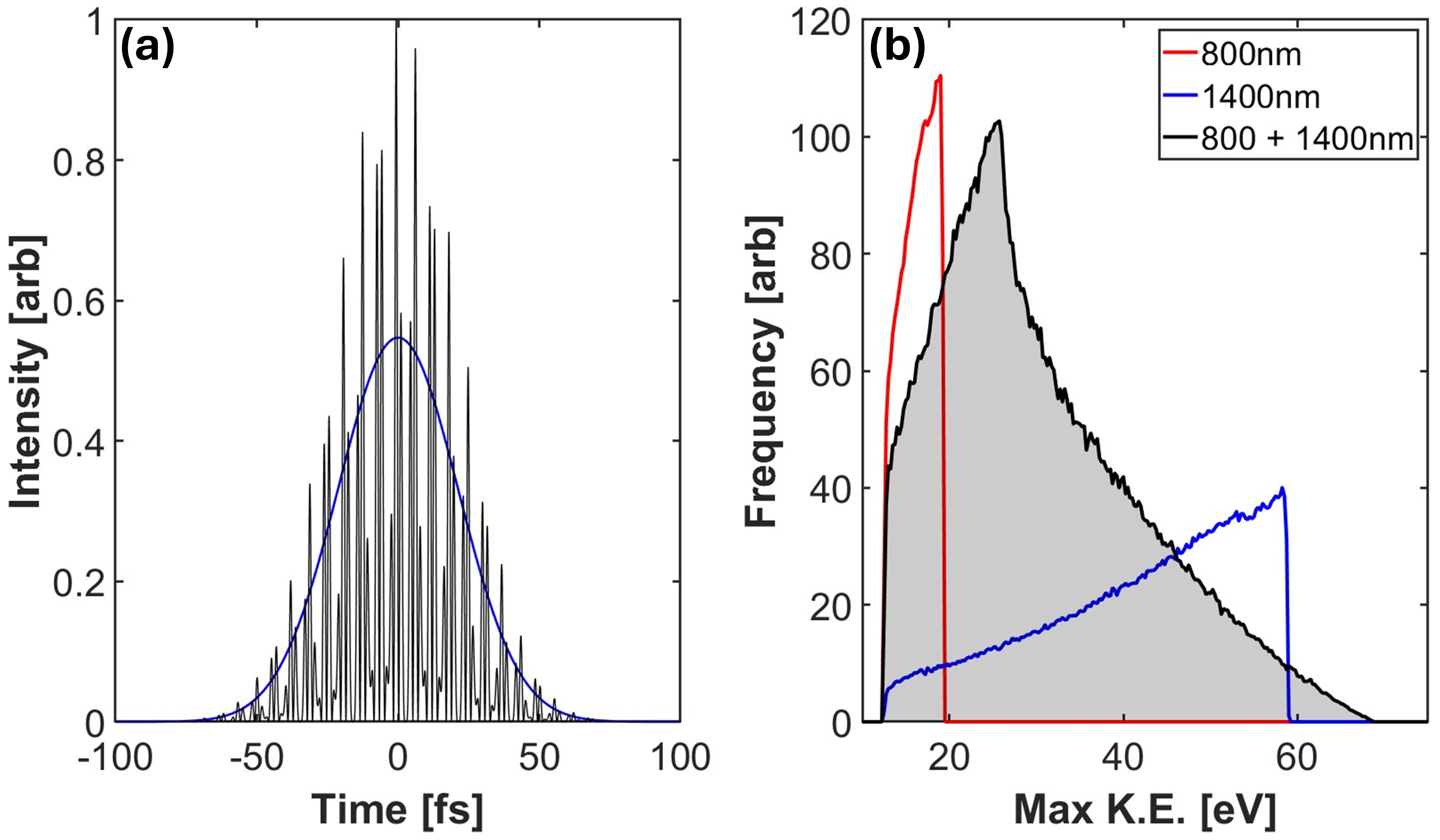}
   \caption{Calculations of the NC fields at 800 and 1400 nm and their resulting electron kinetic energy spectrum upon rescattering. Panel (a) shows the temporal intensity profiles of the combined 800 nm ($2.5\times 10^{13}$ W cm$^{-2}$) and 1400 nm ($7.5\times 10^{13}$ W cm$^{-2}$) fields (black line) with the envelope of a single color field at the total intensity of $1\times 10^{14}$ W cm$^{-2}$. The NC field has a field maximum of 3.7  V/\si{\angstrom} while the single-color fields have a maximum of 2.8 V/\si{\angstrom}. Panel (b) shows the calculated electron kinetic energy distribution upon rescattering for the NC field and for the single color case for 800 nm and 1400 nm. Details about how this panel was calculated are discussed in the Methods section. A version of panel (a) with different relative phases and (b) utilizing the IDS method is shown in Supplementary Figure 1 and 2.}\label{fig1}
 \end{figure}

The single color (800 nm and 1400 nm) vs combined intensity profiles are shown in Figure \ref{fig1}(a). First, we observe that for the same total peak intensity, the 25\%/75\% ratio of 800/1400nm shows significantly larger peaks in the temporal intensity stemming from optical cycles coinciding in phase leading to constructive interference within the pulse envelope, from now on referred to as spikes. Secondly, the intense spikes are spaced further away ($\sim$6 fs) from one another in comparison to a strong monochromatic field, which in the case of 800 nm has a field maximum every 1.33 fs (for the single color fields we only plot the intensity envelope). In the resulting electron kinetic energy spectra following rescattering, shown for the single color cases and the NC case in Figure \ref{fig1}(b), we observe sharp cutoffs for the single-color fields at the theoretically predicted value (3.17 $U_p$). Despite having the same total intensity, however, the combined NC field extended this cutoff beyond the 1400 nm ponderomotive limit to yield even higher electron kinetic energies. The $\lambda^2$ dependence of the ponderomotive energy (Equation \ref{Eq2}) causes the optimal ratio for enhancing the electron's rescattering energy to be biased towards the 1400 nm. We find that the 25\%/75\% intensity ratio for the wavelengths used is optimal for maximizing the electron kinetic energy. Under these NC field conditions, one expects significantly higher internal energy to be deposited into the molecule by the rescattering electron compared to ionization by either single-color fields separately, even when the total laser intensity is matched. In addition, ionizing the molecule by rescattering such energetic electrons is expected to give rise to fragments having a higher appearance energy (AE). This is demonstrated experimentally in the following sections for gas-phase methanol.

As mentioned above, we have employed the IDS method to mitigate the contributions of the lower-intensity regions of the Gaussian laser pulse to the ionization dynamics initiated by the NC field. Wiese et al. computed the intensity distribution of the pulse by taking both the Gaussian spatial and temporal profiles into account.\cite{wiese2019strong} In this 3D configuration, it was shown that the differential volume occupied by the isointensity shell $dI$ around $I$ is given by

 \begin{equation}
   -\frac{\partial V_{3D}}{\partial I}\mbox{d}I \propto \frac{\sqrt{\mbox{ln}\frac{I_0}{I}}}{I} \mbox{d}I,
 \end{equation}
 
which indicates the weight of intensity in the range of 0 to $I_0$ on the interaction of the pulse with the molecular beam. It must be noted that while computing the volume function $V_{3D}$, it is assumed that the radial intensity distribution is independent of the direction of laser beam propagation $z$, since $D<2z_R$ where $D$ and $z_R$ are the molecular beam diameter and the Rayleigh length, respectively. In the present experiments, this interaction region is limited by a 1 mm slit in the extractor plate.

\begin{figure}[H]
    \centering
    \includegraphics[width=0.45\linewidth]{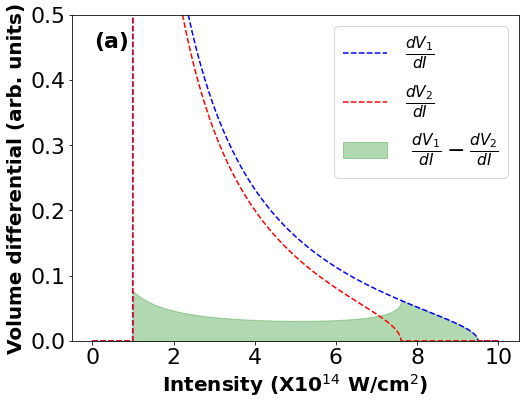}
    \includegraphics[width=0.45\linewidth]{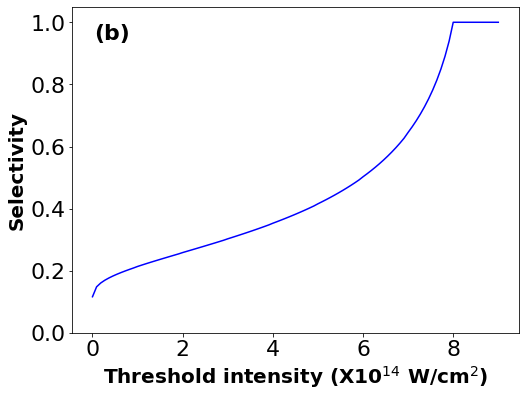}
    \caption{Calculations illustrating the IDS method. Panel (a) plots the volume differential for peak intensity $I_0$ (blue) and $0.8I_0$ (red). The green region corresponds to the difference between the two volume differentials. Panel (b) plots the ion appearance selectivity as a function of increasing threshold intensity. The y-axis represents the ratio of the areas under the difference volume differential ($\Delta \partial V/\partial I$) obtained from species with a high threshold ($8\times10^{14}$ W/cm$^2$) and those with a much lower threshold.}
    \label{fig2}
\end{figure}

The IDS method is conceptually discussed here. Figure \ref{fig2}(a) shows the volume differential for two peak intensities $I_0$ (blue) and $0.8I_0$ (red). Assuming the contribution to the ion yield from regions of the pulse with intensities less than $10^{14}$ W/cm$^2$ is extremely low, we have considered the threshold ionization intensity to be $10^{14}$ W/cm$^2$ in this case, meaning that this intensity is the cutoff for when ionization can begin to occur for the molecule of interest. The green-shaded region refers to the difference between the volume differentials for the two values of peak intensity. The figure shows that the contribution to $\Delta \partial V/\partial I$ by intensities smaller than $0.8I_0$ is highly suppressed. In Figure \ref{fig2} (b), we plot the selectivity of the IDS method in the appearance of ions with increasing AE. The threshold intensity representing the minimum laser intensity at which an ion species appears is plotted along the x-axis. This plot indicates that for ion species having higher threshold intensities, the selectivity of the IDS method is high.

\section{Results and Discussion}
\begin{figure}[H]
    \centering
    \includegraphics[width=.9\linewidth]{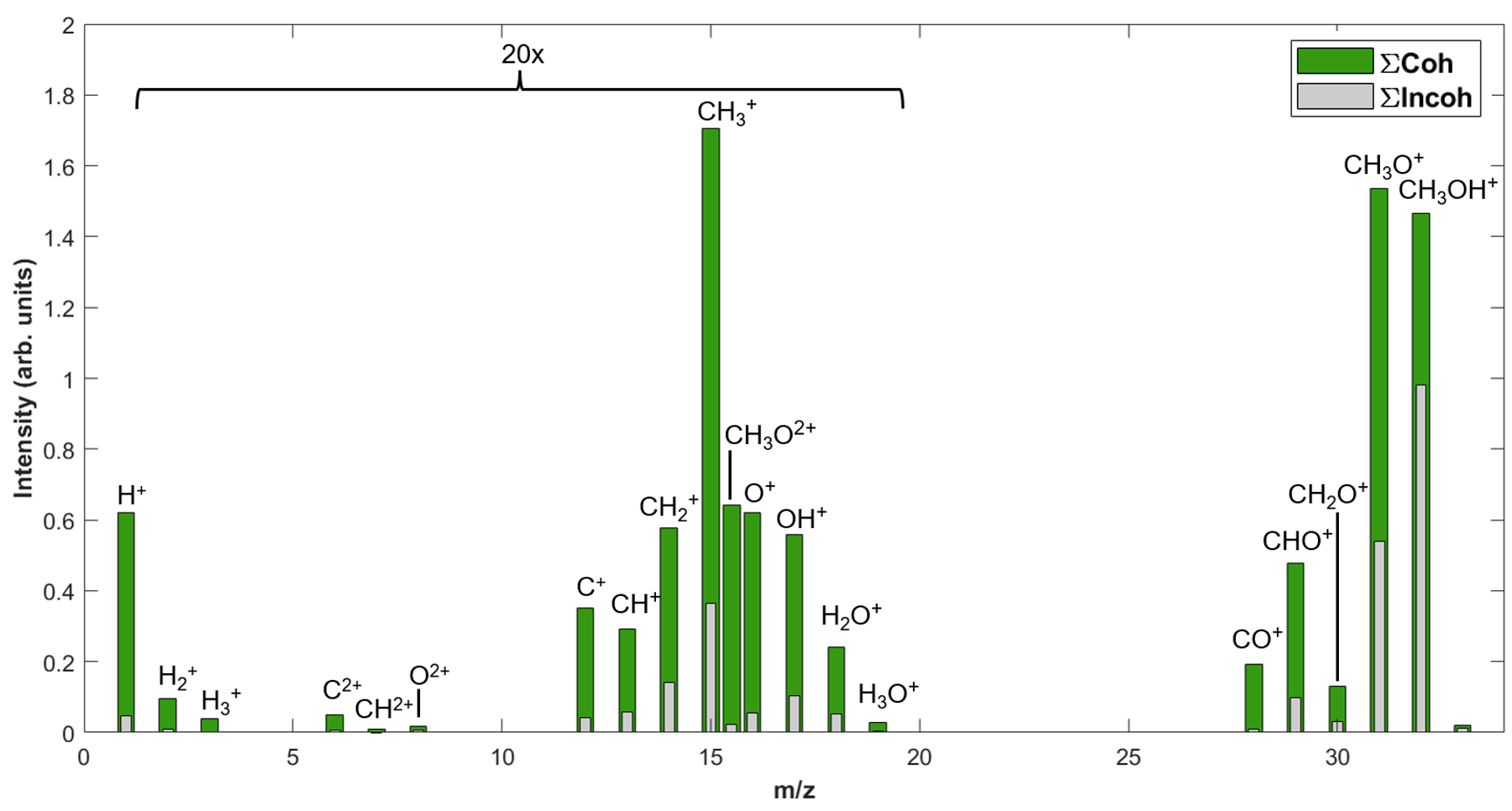}
    \caption{The IDS of methanol with both pulses temporally overlapped (green, coherent sum) and sum of individual pulses (gray, incoherent sum). For the coherent sum, 2.1 x 10$^{14}$ W/cm$^2$ 800 nm, 1.1 x 10$^{14}$ W/cm$^2$ 1400 nm and 1.6 x 10$^{14}$ W/cm$^2$ 800 nm, 8.7 x 10$^{13}$ W/cm$^2$ 1400 nm were used. For the incoherent sum, the same intensities were used individually and resulting spectra summed. Additional results for coherent and single-color fields that produce equivalent total ion yields are shown in Supplementary Figure 3.}
    \label{fig3}
\end{figure}

We ionized methanol with the NC laser fields overlapped in time using two different sets of intensities (2.1 x 10$^{14}$ W/cm$^2$ 800 nm/1.1 x 10$^{14}$ W/cm$^2$ 1400 nm and 1.6 x 10$^{14}$ W/cm$^2$ 800 nm/8.7 x 10$^{13}$ W/cm$^2$ 1400 nm). The resulting IDS obtained by the difference between these two conditions is shown in Figure \ref{fig3}. The spectrum is displayed with the coherent sum ($\sum$Coh) represented in green, indicating the combined effect of the overlapping fields. We compare this result with the IDS obtained from the incoherent sum ($\sum$Incoh) sum of the individual wavelengths at the same powers represented in gray. The analysis reveals an enhanced ion yield from the coherent sum for all ions, contrasting with the lower ion yield observed for the incoherent sum. This enhancement underscores the significant internal energy imparted onto the molecule upon overlapping the two pulses. For the incoherent sum, the spectrum primarily contains the molecular ion (\ce{CH3OH+}) and \ce{CH3O+}, along with a very low yield of \ce{CH2O+}, \ce{CHO+} and \ce{CH3+} (note the 20X region in Figure \ref{fig3}), implying a relatively low internal energy imparted following ionization. From these results, we may infer that the intensities of the individual single-color fields are sufficient to primarily singly ionize the molecules, and fragmentation is insignificant. However, on temporally overlapping the two pulses, causing spikes in the NC electric field, we not only observe a significant enhancement in the yields of the ions observed in the previous case, but also the appearance of smaller fragments mainly originating from dication species such as \ce{H_n+} ($n =$ 1-3), and \ce{CH_n+} ($n=$ 0-3). Despite this, the primary initial charge state of the methanol cation (as determined by the ion yields and identities) in both cases is singly ionized. Ionization with NC fields simply increases the mean internal energy of the methanol molecule by double ionization via electron rescattering. In fact, the enhancement measured is dependent on the AE of the ion, shown in Table \ref{tab1}. This could be explained by the NC field confining energy in selective optical cycles formed by the constructive interference of the fields, causing greater energy to be given to the rescattering electron. This enhances the ion yield by more than a order of magnitude for high-AE fragments. The appearance energy of several fragments of methanol is shown in Table \ref{tab1}.

\begin{table}[H]
    \centering
\caption{Appearance energies and degree of enhancement of combined two-color fields for several key product ions from the ionization of methanol. }
    \begin{tabular}{c|c|cccl}
         Product Ion&m/z&  Appearance Energy (eV) &Other products&Enhancement\\
       \hline  
     
\ce{CH3OH+}&32& 10.85$^{\textcolor{blue}{a}}$ &-&1.5x\\
 \ce{CH3O+}&31&11.67$^{\textcolor{blue}{a}}$ &\ce{H}&2.8x\\
  \ce{CH2O+}&30&10.9$^{\textcolor{blue}{a}}$ &\ce{H2}&4.2x\\
 \ce{CHO+}&29&13.6$^{\textcolor{blue}{a}}$ &\ce{H2}+ \ce{H}&4.8x\\
  \ce{CO+}&28&13.7$^{\textcolor{blue}{a}}$ &2\ce{H2}&19.4x\\
 \ce{CH3OH^2+}/\ce{O^+}&16&32.4$^{\textcolor{blue}{b}}$,30.8$^{\textcolor{blue}{c}}$/12.0$^{\textcolor{blue}{a}}$&-&4.5x\\
 \ce{CH3O^2+}&15.5&33.7$^{\textcolor{blue}{b}}$&-&45.5x\\
\ce{CH3+}&15&13.82$^{\textcolor{blue}{a}}$ &\ce{OH}&4.7x\\
 \ce{CH2+}&14&14.05$^{\textcolor{blue}{a}}$ &\ce{H2O}&4.3x\\
 \ce{CH+}&13&22.31$^{\textcolor{blue}{a}}$&-&5.6x\\
 \ce{C+}&12& 24.4$^{\textcolor{blue}{d}}$ &2\ce{H2} + \ce{O}& 10.1x\\
  \ce{C^2+}&6& 47.9$^{\textcolor{blue}{d}}$&-& $>$60x\\
 \ce{H3+}&3& 31$^{\textcolor{blue}{c}}$ &-& 21.4x\\
\ce{H2+}&2& 26.5$^{\textcolor{blue}{e}}$&\ce{OH} + \ce{C} + \ce{H}& 16.2x\\
 \ce{H+}&1& 21.5$^{\textcolor{blue}{e}}$&\ce{H2O} + \ce{CH}& 15.4x\\
 \hline
 \end{tabular}
 \vspace{6pt} 
    \begin{minipage}{\textwidth}
       Data obtained from \textcolor{blue}{a}. NIST\cite{NIST}
        \textcolor{blue}{b}. Douglas et al.\cite{douglas2009studies}
        \textcolor{blue}{c}. Eland et al.\cite{Brown1992}
        \textcolor{blue}{d}. Atomic ionization energy\cite{AIE}
        \textcolor{blue}{e}. Burton et al.\cite{Burton1992}
    \end{minipage}
    \label{tab1}
\end{table}

We can relate the AE to the degree of enhancement as follows: According to the previous discussion, the main difference between $\sum$Coh and $\sum$Incoh is the increase in the rescattering energy of the electron. Therefore, it is expected that ions requiring more energy for their creation show the greatest difference. Indeed, ions with higher AE exhibit the largest enhancement in energetic order. Specifically, C$^+$ $<$ H$^+$ $<$ H$_2^+$ $<$ H$_3^+$ $<$ CH$_3$O$^{2+}$ $<$ C$^{2+}$ (see Table \ref{tab1}) align with the theoretical predictions. In the case of C$^{2+}$, the $\sum$Incoh mass spectrum showed no signal at this m/z, therefore the number given is based on the experimental noise floor. On the other hand, ions with low AE (with the exception of \ce{CO+}) show poor enhancement. One possible explanation for the greater-than-expected enhancement observed for \ce{CO+} is that this ion could also be created through a high-energy mechanism, such as the fragmentation of doubly charged methanol (\ce{CH3OH^2+}), as depicted in the following reaction.\cite{douglas2009studies}

\begin{center}
CH$_3$OH$^{2+}$ $\rightarrow$ CH$_2$O$^+$ + H$_2^+$\\
CH$_2$O$^+$ $\rightarrow$ CO$^+$ + H$_2$
\end{center}

\begin{figure}[H]
    \centering
    \includegraphics[width=0.9\linewidth]{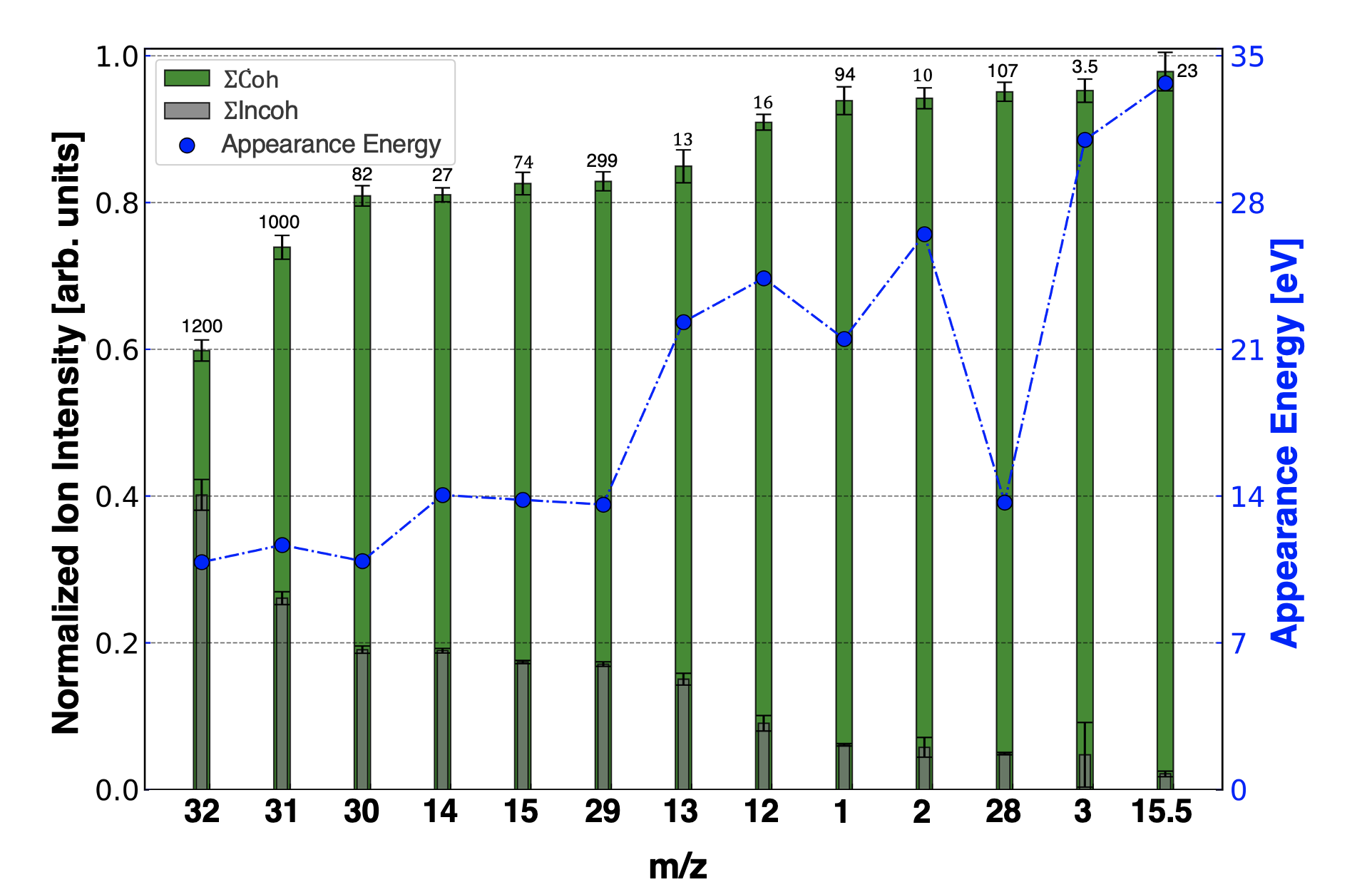}
    \caption{
    Bar graph of selected ions from IDS that have been normalized such that the incoherent sum (grey) plus the coherent sum (green) equals one. The number on top of each bar indicates the factor by which the raw ion yield has been divided to normalize it. The error bars display $\pm$ 1 standard deviation of the ion yield from the 64 scans. The blue circles in each bar, connected by a dashed line, correspond to the appearance energy of each fragment ion. 
    }
    \label{fig4}
\end{figure}

Considering that the threshold of the first fragmentation requires approximately 31.5 eV\cite{Brown1992} and that an excess of internal energy of CH$_2$O$^+$ is necessary to produce CO$^+$, it is reasonable to observe the enhancement of this ion. We expect high-energy mechanisms to explain other departures from the trend in Figure \ref{fig4}, such as H$^+$. By normalizing the experimental results such that the combined yield of coherent and incoherent sums equals one, we achieve a clearer representation of the enhancement, illustrated in Figure \ref{fig4}. Upon arranging the results from minimum to maximum enhancement, a correlation emerges between enhancement and the appearance energy of each fragment ion. Notably, CO$^+$ stands as an outlier, as previously noted. This correlation aligns with insights from the analysis depicted in Figure \ref{fig2}b, where the IDS method's selectivity correlates with the threshold intensity needed to detect a specific ion.

\section{Conclusions}
The effect of combining non-commensurate laser field ionization with the IDS method to study the dissociative ionization of polyatomic molecules is demonstrated for methanol. We observed over an order of magnitude increase in specific ionization pathways when using NC fields when compared to the incoherent sum of fields, an enhancement correlated with their respective appearance energies. This effect is attributed to sparse instances of constructive interference between the non-commensurate optical cycles, leading to a greater kinetic energy gain of the rescattered electron. These findings confirm what had been observed for high harmonic generation, namely the generation of electrons with higher kinetic energy achieved by the coherent sum of NC fields, determined by the presence of high appearance-energy fragments. These results allow for the study of  chemistry induced by electrons with chemically relevant energies, in particular, when the energy of individual fields is insufficient. Additionally, the use of NC fields restricts ionization to individual instances where constructive interference occurs between the two optical fields. The combination of NC fields with IDS minimizes the contribution from multiphoton ionization. Future experiments in our laboratory will explore other pulse parameters, such as different wavelengths, polarizations, and pulse durations, to control electron-initiated reactions.

\section{Methods}
Briefly, the experimental setup is described below. The two-color field comprising non-commensurate frequencies was obtained by overlapping parallel polarized pulses having central wavelengths of 800 nm and 1400 nm. A Ti:sapphire laser (ASTRELLA, Coherent Inc.) producing 800 nm, 5 mJ, 1 kHz pulses was split 50/50 using a beam splitter, a part of which was sent into an optical parametric amplifier (TOPAS Prime, Light Conversion) to produce tunable laser pulses from 260 nm to 2600 nm. The 1400 nm beam was expanded using a pair of lenses to ensure both beams had the same focal length and Rayleigh lengths when collinearly combined and focused by a single lens. The 1400 nm and 800 nm pulses were measured to have a pulse duration of 53 fs and 168 fs, respectively. The beams were collinearly combined and focused into a Wiley-McLaren time-of-flight (TOF) mass spectrometer. The intensities of the pulses were controlled using variable attenuators. The two-color pulses were focused in the ionization region of the TOF spectrometer with an achromatic doublet lens (focal length = 200 mm).

At the average intensities maintained during the experiment ($2.1\times10^{14}$ W/cm$^2$ for 800 nm and $1.1\times10^{14}$ W/cm$^2$ for 1400 nm pulses). These intensities were determined in situ by measuring the ratio of Ar$^{2+}$/Ar$^+$.\cite{guo1998single} The Keldysh parameter for the strong-field ionization of methanol (ionization potential = 10.85 eV) was found to be 0.38 and 0.157, respectively. This indicates that the ionization by the two-color pulses occurs predominately via tunnel ionization. Furthermore, the NC mass spectrum of methanol is in good agreement with the experimental tunnel ionization mass spectrum of methanol in Rajgara, et al. where an electron rescattering mechanism was concluded.\cite{rajgara2003electron} Dry methanol was effused into the TOF chamber using a needle valve. The baseline chamber pressure was $9\times10^{-8}$ Torr, while the pressure was maintained at $2\times10^{-6}$ Torr, corresponding to a density of $7 \times 10^{10}$ molecules/cm\textsuperscript{3},  throughout the experiment. The ion TOF signals were detected using a microchannel plate (MCP, RM Jordan) detector and digitized using an oscilloscope (LeCroy WaveRunner 610Zi, 1 GHz). Each scan was comprised of 300 laser shots and was repeated 64 times.

The simulations in Figure \ref{fig1}(b) result from a trajectory-based classical mechanics simulation of the free electron in the optical field. For each trajectory, the electron is born at a random point within the NC field (where the field is given a random phase) and evolved according to classical mechanics under the action of the electric field of the pulse. If the electron returns within a certain radius of the nucleus (a radius given by the scattering cross section of the electron at its instantaneous kinetic energy), the trajectory is labelled with that kinetic energy. This result is then scaled according to the ADK probability that the electron was birthed at a starting point given the pulse’s electric field at that time. This process was done for 5000 random NC phases and each phase was ran with 5000 random electron birth points. The histogram of the resulting rescattering energies is shown in Figure \ref{fig1}(b). Both NC fields were parallel polarized, meaning no elaborate electron orbits occurred and the combined fields only acted to increase the electron kinetic energy ceiling.

\section{Acknowledgements}
This material is based upon work supported by the U.S. Department of Energy, Office of Science, Office of Basic Energy Sciences, Atomic, Molecular, and Optical Sciences Program under SISGER DE-SC0002325. Additional funding to support the contributions from RD and SK comes from the Air Force Office of Scientific Research under award number FA9550-21-1-0428.

\bibliography{references}

\providecommand{\latin}[1]{#1}
\makeatletter
\providecommand{\doi}
  {\begingroup\let\do\@makeother\dospecials
  \catcode`\{=1 \catcode`\}=2 \doi@aux}
\providecommand{\doi@aux}[1]{\endgroup\texttt{#1}}
\makeatother
\providecommand*\mcitethebibliography{\thebibliography}
\csname @ifundefined\endcsname{endmcitethebibliography}  {\let\endmcitethebibliography\endthebibliography}{}
\begin{mcitethebibliography}{53}
\providecommand*\natexlab[1]{#1}
\providecommand*\mciteSetBstSublistMode[1]{}
\providecommand*\mciteSetBstMaxWidthForm[2]{}
\providecommand*\mciteBstWouldAddEndPuncttrue
  {\def\EndOfBibitem{\unskip.}}
\providecommand*\mciteBstWouldAddEndPunctfalse
  {\let\EndOfBibitem\relax}
\providecommand*\mciteSetBstMidEndSepPunct[3]{}
\providecommand*\mciteSetBstSublistLabelBeginEnd[3]{}
\providecommand*\EndOfBibitem{}
\mciteSetBstSublistMode{f}
\mciteSetBstMaxWidthForm{subitem}{(\alph{mcitesubitemcount})}
\mciteSetBstSublistLabelBeginEnd
  {\mcitemaxwidthsubitemform\space}
  {\relax}
  {\relax}

\bibitem[Corkum(1993)]{corkum1993plasma}
Corkum,~P.~B. Plasma perspective on strong field multiphoton ionization. \emph{Physical review letters} \textbf{1993}, \emph{71}, 1994\relax
\mciteBstWouldAddEndPuncttrue
\mciteSetBstMidEndSepPunct{\mcitedefaultmidpunct}
{\mcitedefaultendpunct}{\mcitedefaultseppunct}\relax
\EndOfBibitem
\bibitem[Niikura \latin{et~al.}(2002)Niikura, L{\'e}gar{\'e}, Hasbani, Bandrauk, Ivanov, Villeneuve, and Corkum]{niikura2002sub}
Niikura,~H.; L{\'e}gar{\'e},~F.; Hasbani,~R.; Bandrauk,~A.; Ivanov,~M.~Y.; Villeneuve,~D.; Corkum,~P. Sub-laser-cycle electron pulses for probing molecular dynamics. \emph{Nature} \textbf{2002}, \emph{417}, 917--922\relax
\mciteBstWouldAddEndPuncttrue
\mciteSetBstMidEndSepPunct{\mcitedefaultmidpunct}
{\mcitedefaultendpunct}{\mcitedefaultseppunct}\relax
\EndOfBibitem
\bibitem[Li \latin{et~al.}(2020)Li, Sierra-Costa, Michie, Ben-Itzhak, and Dantus]{li2020control}
Li,~S.; Sierra-Costa,~D.; Michie,~M.~J.; Ben-Itzhak,~I.; Dantus,~M. Control of electron recollision and molecular nonsequential double ionization. \emph{Communications Physics} \textbf{2020}, \emph{3}, 35\relax
\mciteBstWouldAddEndPuncttrue
\mciteSetBstMidEndSepPunct{\mcitedefaultmidpunct}
{\mcitedefaultendpunct}{\mcitedefaultseppunct}\relax
\EndOfBibitem
\bibitem[Dantus(2024)]{dantus2024tracking}
Dantus,~M. Tracking Molecular Fragmentation in Electron--Ionization Mass Spectrometry with Ultrafast Time Resolution. \emph{Accounts of Chemical Research} \textbf{2024}, 033003\relax
\mciteBstWouldAddEndPuncttrue
\mciteSetBstMidEndSepPunct{\mcitedefaultmidpunct}
{\mcitedefaultendpunct}{\mcitedefaultseppunct}\relax
\EndOfBibitem
\bibitem[Dantus(2024)]{dantussci2024}
Dantus,~M. Ultrafast Studies of Elusive Chemical Reactions in the Gas Phase. \emph{Science Review} \textbf{2024}, \emph{385}, eadk1833\relax
\mciteBstWouldAddEndPuncttrue
\mciteSetBstMidEndSepPunct{\mcitedefaultmidpunct}
{\mcitedefaultendpunct}{\mcitedefaultseppunct}\relax
\EndOfBibitem
\bibitem[Mics \latin{et~al.}(2005)Mics, Kadlec, Ku{\v{z}}el, Jungwirth, Bradforth, and Apkarian]{mics2005nonresonant}
Mics,~Z.; Kadlec,~F.; Ku{\v{z}}el,~P.; Jungwirth,~P.; Bradforth,~S.~E.; Apkarian,~V. Nonresonant ionization of oxygen molecules by femtosecond pulses: Plasma dynamics studied by time-resolved terahertz spectroscopy. \emph{The Journal of chemical physics} \textbf{2005}, \emph{123}, 104310\relax
\mciteBstWouldAddEndPuncttrue
\mciteSetBstMidEndSepPunct{\mcitedefaultmidpunct}
{\mcitedefaultendpunct}{\mcitedefaultseppunct}\relax
\EndOfBibitem
\bibitem[Nairat \latin{et~al.}(2016)Nairat, Lozovoy, and Dantus]{nairat2016order}
Nairat,~M.; Lozovoy,~V.~V.; Dantus,~M. Order of magnitude dissociative ionization enhancement observed for pulses with high order dispersion. \emph{The Journal of Physical Chemistry A} \textbf{2016}, \emph{120}, 8529--8536\relax
\mciteBstWouldAddEndPuncttrue
\mciteSetBstMidEndSepPunct{\mcitedefaultmidpunct}
{\mcitedefaultendpunct}{\mcitedefaultseppunct}\relax
\EndOfBibitem
\bibitem[Bandulet \latin{et~al.}(2010)Bandulet, Comtois, Bisson, Fleischer, P{\'e}pin, Kieffer, Corkum, and Villeneuve]{bandulet2010gating}
Bandulet,~H.-C.; Comtois,~D.; Bisson,~E.; Fleischer,~A.; P{\'e}pin,~H.; Kieffer,~J.-C.; Corkum,~P.~B.; Villeneuve,~D.~M. Gating attosecond pulse train generation using multicolor laser fields. \emph{Physical Review A} \textbf{2010}, \emph{81}, 013803\relax
\mciteBstWouldAddEndPuncttrue
\mciteSetBstMidEndSepPunct{\mcitedefaultmidpunct}
{\mcitedefaultendpunct}{\mcitedefaultseppunct}\relax
\EndOfBibitem
\bibitem[Bruner \latin{et~al.}(2018)Bruner, Kr{\"u}ger, Pedatzur, Orenstein, Azoury, and Dudovich]{bruner2018robust}
Bruner,~B.~D.; Kr{\"u}ger,~M.; Pedatzur,~O.; Orenstein,~G.; Azoury,~D.; Dudovich,~N. Robust enhancement of high harmonic generation via attosecond control of ionization. \emph{Optics express} \textbf{2018}, \emph{26}, 9310--9322\relax
\mciteBstWouldAddEndPuncttrue
\mciteSetBstMidEndSepPunct{\mcitedefaultmidpunct}
{\mcitedefaultendpunct}{\mcitedefaultseppunct}\relax
\EndOfBibitem
\bibitem[Bruner \latin{et~al.}(2021)Bruner, Narovlansky-Uzan, Arusi-Parpar, Orenstein, Shonfeld, and Dudovich]{bruner2021control}
Bruner,~B.~D.; Narovlansky-Uzan,~A.~J.; Arusi-Parpar,~T.; Orenstein,~G.; Shonfeld,~A.; Dudovich,~N. Control and enhancement of multiband high harmonic generation by synthesized laser fields. \emph{Journal of Physics B: Atomic, Molecular and Optical Physics} \textbf{2021}, \emph{54}, 154001\relax
\mciteBstWouldAddEndPuncttrue
\mciteSetBstMidEndSepPunct{\mcitedefaultmidpunct}
{\mcitedefaultendpunct}{\mcitedefaultseppunct}\relax
\EndOfBibitem
\bibitem[Posthumus(2004)]{posthumus2004dynamics}
Posthumus,~J. The dynamics of small molecules in intense laser fields. \emph{Reports on Progress in Physics} \textbf{2004}, \emph{67}, 623\relax
\mciteBstWouldAddEndPuncttrue
\mciteSetBstMidEndSepPunct{\mcitedefaultmidpunct}
{\mcitedefaultendpunct}{\mcitedefaultseppunct}\relax
\EndOfBibitem
\bibitem[Wang \latin{et~al.}(2005)Wang, Sayler, Carnes, Esry, and Ben-Itzhak]{wang2005disentangling}
Wang,~P.; Sayler,~A.~M.; Carnes,~K.~D.; Esry,~B.~D.; Ben-Itzhak,~I. Disentangling the volume effect through intensity-difference spectra:? application to laser-induced dissociation of H2+. \emph{Optics letters} \textbf{2005}, \emph{30}, 664--666\relax
\mciteBstWouldAddEndPuncttrue
\mciteSetBstMidEndSepPunct{\mcitedefaultmidpunct}
{\mcitedefaultendpunct}{\mcitedefaultseppunct}\relax
\EndOfBibitem
\bibitem[Wiese \latin{et~al.}(2019)Wiese, Olivieri, Trabattoni, Trippel, and K{\"u}pper]{wiese2019strong}
Wiese,~J.; Olivieri,~J.-F.; Trabattoni,~A.; Trippel,~S.; K{\"u}pper,~J. Strong-field photoelectron momentum imaging of OCS at finely resolved incident intensities. \emph{New Journal of Physics} \textbf{2019}, \emph{21}, 083011\relax
\mciteBstWouldAddEndPuncttrue
\mciteSetBstMidEndSepPunct{\mcitedefaultmidpunct}
{\mcitedefaultendpunct}{\mcitedefaultseppunct}\relax
\EndOfBibitem
\bibitem[Muller \latin{et~al.}(1990)Muller, Bucksbaum, Schumacher, and Zavriyev]{muller1990above}
Muller,~H.; Bucksbaum,~P.; Schumacher,~D.; Zavriyev,~A. Above-threshold ionisation with a two-colour laser field. \emph{Journal of Physics B: Atomic, Molecular and Optical Physics} \textbf{1990}, \emph{23}, 2761\relax
\mciteBstWouldAddEndPuncttrue
\mciteSetBstMidEndSepPunct{\mcitedefaultmidpunct}
{\mcitedefaultendpunct}{\mcitedefaultseppunct}\relax
\EndOfBibitem
\bibitem[Schafer and Kulander(1992)Schafer, and Kulander]{schafer1992phase}
Schafer,~K.~J.; Kulander,~K.~C. Phase-dependent effects in multiphoton ionization induced by a laser field and its second harmonic. \emph{Physical Review A} \textbf{1992}, \emph{45}, 8026\relax
\mciteBstWouldAddEndPuncttrue
\mciteSetBstMidEndSepPunct{\mcitedefaultmidpunct}
{\mcitedefaultendpunct}{\mcitedefaultseppunct}\relax
\EndOfBibitem
\bibitem[Schumacher \latin{et~al.}(1994)Schumacher, Weihe, Muller, and Bucksbaum]{schumacher1994phase}
Schumacher,~D.; Weihe,~F.; Muller,~H.; Bucksbaum,~P. Phase dependence of intense field ionization: a study using two colors. \emph{Physical review letters} \textbf{1994}, \emph{73}, 1344\relax
\mciteBstWouldAddEndPuncttrue
\mciteSetBstMidEndSepPunct{\mcitedefaultmidpunct}
{\mcitedefaultendpunct}{\mcitedefaultseppunct}\relax
\EndOfBibitem
\bibitem[Fang \latin{et~al.}(2019)Fang, He, Han, Ge, Yu, Ma, Deng, and Liu]{fang2019strong}
Fang,~Y.; He,~C.; Han,~M.; Ge,~P.; Yu,~X.; Ma,~X.; Deng,~Y.; Liu,~Y. Strong-field ionization of Ar atoms with a 45° cross-linearly-polarized two-color laser field. \emph{Physical Review A} \textbf{2019}, \emph{100}, 013414\relax
\mciteBstWouldAddEndPuncttrue
\mciteSetBstMidEndSepPunct{\mcitedefaultmidpunct}
{\mcitedefaultendpunct}{\mcitedefaultseppunct}\relax
\EndOfBibitem
\bibitem[Liu \latin{et~al.}(2018)Liu, Han, Ge, He, Gong, and Liu]{liu2018strong}
Liu,~M.-M.; Han,~M.; Ge,~P.; He,~C.; Gong,~Q.; Liu,~Y. Strong-field ionization of diatomic molecules in orthogonally polarized two-color fields. \emph{Physical Review A} \textbf{2018}, \emph{97}, 063416\relax
\mciteBstWouldAddEndPuncttrue
\mciteSetBstMidEndSepPunct{\mcitedefaultmidpunct}
{\mcitedefaultendpunct}{\mcitedefaultseppunct}\relax
\EndOfBibitem
\bibitem[Kotsina \latin{et~al.}(2015)Kotsina, Kaziannis, and Kosmidis]{kotsina2015phase}
Kotsina,~N.; Kaziannis,~S.; Kosmidis,~C. Phase dependence of OD+, HOD+, and H3+ ions released from the deuterated dication of methanol under $\omega$/2$\omega$ laser field irradiation. \emph{International Journal of Mass Spectrometry} \textbf{2015}, \emph{380}, 34--39\relax
\mciteBstWouldAddEndPuncttrue
\mciteSetBstMidEndSepPunct{\mcitedefaultmidpunct}
{\mcitedefaultendpunct}{\mcitedefaultseppunct}\relax
\EndOfBibitem
\bibitem[Kechaoglou \latin{et~al.}(2019)Kechaoglou, Kaziannis, and Kosmidis]{kechaoglou2019controlling}
Kechaoglou,~E.; Kaziannis,~S.; Kosmidis,~C. Controlling intramolecular hydrogen migration by asymmetric laser fields: the water case. \emph{Physical Chemistry Chemical Physics} \textbf{2019}, \emph{21}, 11259--11265\relax
\mciteBstWouldAddEndPuncttrue
\mciteSetBstMidEndSepPunct{\mcitedefaultmidpunct}
{\mcitedefaultendpunct}{\mcitedefaultseppunct}\relax
\EndOfBibitem
\bibitem[Walt \latin{et~al.}(2015)Walt, Bhargava~Ram, Von~Conta, Tolstikhin, Madsen, Jensen, and Wörner]{walt2015role}
Walt,~S.~G.; Bhargava~Ram,~N.; Von~Conta,~A.; Tolstikhin,~O.~I.; Madsen,~L.~B.; Jensen,~F.; Wörner,~H.~J. Role of multi-electron effects in the asymmetry of strong-field ionization and fragmentation of polar molecules: the methyl halide series. \emph{The Journal of Physical Chemistry A} \textbf{2015}, \emph{119}, 11772--11782\relax
\mciteBstWouldAddEndPuncttrue
\mciteSetBstMidEndSepPunct{\mcitedefaultmidpunct}
{\mcitedefaultendpunct}{\mcitedefaultseppunct}\relax
\EndOfBibitem
\bibitem[Luo \latin{et~al.}(2017)Luo, Li, Xie, Zhang, Xu, Li, Zhou, Lan, and Lu]{luo2017angular}
Luo,~S.; Li,~M.; Xie,~H.; Zhang,~P.; Xu,~S.; Li,~Y.; Zhou,~Y.; Lan,~P.; Lu,~P. Angular-dependent asymmetries of above-threshold ionization in a two-color laser field. \emph{Physical Review A} \textbf{2017}, \emph{96}, 023417\relax
\mciteBstWouldAddEndPuncttrue
\mciteSetBstMidEndSepPunct{\mcitedefaultmidpunct}
{\mcitedefaultendpunct}{\mcitedefaultseppunct}\relax
\EndOfBibitem
\bibitem[Madhusudhan \latin{et~al.}(2022)Madhusudhan, Das, Bhardwaj, KM, Nimma, Soumyashree, and Kushawaha]{madhusudhan2022strong}
Madhusudhan,~P.; Das,~R.; Bhardwaj,~P.; KM,~M.~S.; Nimma,~V.; Soumyashree,~S.; Kushawaha,~R.~K. Strong-field ionization of N2 and CO molecules using two-color laser field. \emph{Journal of Physics B: Atomic, Molecular and Optical Physics} \textbf{2022}, \emph{55}, 234001\relax
\mciteBstWouldAddEndPuncttrue
\mciteSetBstMidEndSepPunct{\mcitedefaultmidpunct}
{\mcitedefaultendpunct}{\mcitedefaultseppunct}\relax
\EndOfBibitem
\bibitem[Betsch \latin{et~al.}(2010)Betsch, Pinkham, and Jones]{betsch2010directional}
Betsch,~K.; Pinkham,~D.; Jones,~R. Directional emission of multiply charged ions during dissociative ionization in asymmetric two-color laser fields. \emph{Physical review letters} \textbf{2010}, \emph{105}, 223002\relax
\mciteBstWouldAddEndPuncttrue
\mciteSetBstMidEndSepPunct{\mcitedefaultmidpunct}
{\mcitedefaultendpunct}{\mcitedefaultseppunct}\relax
\EndOfBibitem
\bibitem[Kaziannis \latin{et~al.}(2014)Kaziannis, Kotsina, and Kosmidis]{kaziannis2014interaction}
Kaziannis,~S.; Kotsina,~N.; Kosmidis,~C. Interaction of toluene with two-color asymmetric laser fields: Controlling the directional emission of molecular hydrogen fragments. \emph{The Journal of Chemical Physics} \textbf{2014}, \emph{141}, 104319\relax
\mciteBstWouldAddEndPuncttrue
\mciteSetBstMidEndSepPunct{\mcitedefaultmidpunct}
{\mcitedefaultendpunct}{\mcitedefaultseppunct}\relax
\EndOfBibitem
\bibitem[Song \latin{et~al.}(2015)Song, Gong, Ji, Lin, Pan, Ding, Zeng, and Wu]{song2015directional}
Song,~Q.; Gong,~X.; Ji,~Q.; Lin,~K.; Pan,~H.; Ding,~J.; Zeng,~H.; Wu,~J. Directional deprotonation ionization of acetylene in asymmetric two-color laser fields. \emph{Journal of Physics B: Atomic, Molecular and Optical Physics} \textbf{2015}, \emph{48}, 094007\relax
\mciteBstWouldAddEndPuncttrue
\mciteSetBstMidEndSepPunct{\mcitedefaultmidpunct}
{\mcitedefaultendpunct}{\mcitedefaultseppunct}\relax
\EndOfBibitem
\bibitem[Watanabe \latin{et~al.}(1994)Watanabe, Kondo, Nabekawa, Sagisaka, and Kobayashi]{watanabe1994two}
Watanabe,~S.; Kondo,~K.; Nabekawa,~Y.; Sagisaka,~A.; Kobayashi,~Y. Two-color phase control in tunneling ionization and harmonic generation by a strong laser field and its third harmonic. \emph{Physical review letters} \textbf{1994}, \emph{73}, 2692\relax
\mciteBstWouldAddEndPuncttrue
\mciteSetBstMidEndSepPunct{\mcitedefaultmidpunct}
{\mcitedefaultendpunct}{\mcitedefaultseppunct}\relax
\EndOfBibitem
\bibitem[Kondo \latin{et~al.}(1996)Kondo, Kobayashi, Sagisaka, Nabekawa, and Watanabe]{kondo1996tunneling}
Kondo,~K.; Kobayashi,~Y.; Sagisaka,~A.; Nabekawa,~Y.; Watanabe,~S. Tunneling ionization and harmonic generation in two-color fields. \emph{JOSA B} \textbf{1996}, \emph{13}, 424--429\relax
\mciteBstWouldAddEndPuncttrue
\mciteSetBstMidEndSepPunct{\mcitedefaultmidpunct}
{\mcitedefaultendpunct}{\mcitedefaultseppunct}\relax
\EndOfBibitem
\bibitem[Cormier and Lewenstein(2000)Cormier, and Lewenstein]{cormier2000optimizing}
Cormier,~E.; Lewenstein,~M. Optimizing the efficiency in high order harmonic generation optimization by two-color fields. \emph{The European Physical Journal D-Atomic, Molecular, Optical and Plasma Physics} \textbf{2000}, \emph{12}, 227--233\relax
\mciteBstWouldAddEndPuncttrue
\mciteSetBstMidEndSepPunct{\mcitedefaultmidpunct}
{\mcitedefaultendpunct}{\mcitedefaultseppunct}\relax
\EndOfBibitem
\bibitem[Zhai \latin{et~al.}(2020)Zhai, Shao, Lan, Wang, Zhang, Yuan, Njoroge, He, and Lu]{zhai2020ellipticity}
Zhai,~C.; Shao,~R.; Lan,~P.; Wang,~B.; Zhang,~Y.; Yuan,~H.; Njoroge,~S.~M.; He,~L.; Lu,~P. Ellipticity control of high-order harmonic generation with nearly orthogonal two-color laser fields. \emph{Physical Review A} \textbf{2020}, \emph{101}, 053407\relax
\mciteBstWouldAddEndPuncttrue
\mciteSetBstMidEndSepPunct{\mcitedefaultmidpunct}
{\mcitedefaultendpunct}{\mcitedefaultseppunct}\relax
\EndOfBibitem
\bibitem[Siegel \latin{et~al.}(2010)Siegel, Torres, Hoffmann, Brugnera, Procino, Za{\"\i}r, Underwood, Springate, Turcu, Chipperfield, \latin{et~al.} others]{siegel2010high}
Siegel,~T.; Torres,~R.; Hoffmann,~D.; Brugnera,~L.; Procino,~I.; Za{\"\i}r,~A.; Underwood,~J.~G.; Springate,~E.; Turcu,~I.; Chipperfield,~L.; others High harmonic emission from a superposition of multiple unrelated frequency fields. \emph{Optics Express} \textbf{2010}, \emph{18}, 6853--6862\relax
\mciteBstWouldAddEndPuncttrue
\mciteSetBstMidEndSepPunct{\mcitedefaultmidpunct}
{\mcitedefaultendpunct}{\mcitedefaultseppunct}\relax
\EndOfBibitem
\bibitem[Takahashi \latin{et~al.}(2010)Takahashi, Lan, M{\"u}cke, Nabekawa, and Midorikawa]{takahashi2010infrared}
Takahashi,~E.~J.; Lan,~P.; M{\"u}cke,~O.~D.; Nabekawa,~Y.; Midorikawa,~K. Infrared two-color multicycle laser field synthesis for generating an intense attosecond pulse. \emph{Physical review letters} \textbf{2010}, \emph{104}, 233901\relax
\mciteBstWouldAddEndPuncttrue
\mciteSetBstMidEndSepPunct{\mcitedefaultmidpunct}
{\mcitedefaultendpunct}{\mcitedefaultseppunct}\relax
\EndOfBibitem
\bibitem[Lan \latin{et~al.}(2010)Lan, Takahashi, and Midorikawa]{lan2010optimization}
Lan,~P.; Takahashi,~E.~J.; Midorikawa,~K. Optimization of infrared two-color multicycle field synthesis for intense-isolated-attosecond-pulse generation. \emph{Physical Review A} \textbf{2010}, \emph{82}, 053413\relax
\mciteBstWouldAddEndPuncttrue
\mciteSetBstMidEndSepPunct{\mcitedefaultmidpunct}
{\mcitedefaultendpunct}{\mcitedefaultseppunct}\relax
\EndOfBibitem
\bibitem[{Charles S. Cummings, and Walker Bleakney}(1940)]{Charles1940}
{Charles S. Cummings, and Walker Bleakney} {Products of ionization by electron impact in methyl and ethyl alcohol}. \emph{Physical Review Journals} \textbf{1940}, \emph{58}, 787--792\relax
\mciteBstWouldAddEndPuncttrue
\mciteSetBstMidEndSepPunct{\mcitedefaultmidpunct}
{\mcitedefaultendpunct}{\mcitedefaultseppunct}\relax
\EndOfBibitem
\bibitem[{ C Szmytkowski and A M Krzysztofowicz}(1995)]{Szmytkowski1995}
{ C Szmytkowski and A M Krzysztofowicz} {Electron scattering from isoelectronic, Ne=18, CH$_3$X molecules (X=F, OH, NH$_2$ and CH$_3$)}. \emph{Journal of Physics B: Atomic, Molecular and Optical Physics} \textbf{1995}, \emph{28}, 015201\relax
\mciteBstWouldAddEndPuncttrue
\mciteSetBstMidEndSepPunct{\mcitedefaultmidpunct}
{\mcitedefaultendpunct}{\mcitedefaultseppunct}\relax
\EndOfBibitem
\bibitem[{Satyendra Pal}(2004)]{Satyendra2004}
{Satyendra Pal} {Determination of single differential and partial cross-sections for the production of cations in electron–methanol collision}. \emph{Chemical Physics} \textbf{2004}, \emph{302}, 119–124\relax
\mciteBstWouldAddEndPuncttrue
\mciteSetBstMidEndSepPunct{\mcitedefaultmidpunct}
{\mcitedefaultendpunct}{\mcitedefaultseppunct}\relax
\EndOfBibitem
\bibitem[{D G M Silva, T Tejo, J Muse, D Romero, M A Khakoo and M C A Lopes}(2010)]{Silva2010}
{D G M Silva, T Tejo, J Muse, D Romero, M A Khakoo and M C A Lopes} {Total electron scattering cross sections for methanol and ethanol at intermediate energies}. \emph{Journal of Physics B: Atomic, Molecular and Optical Physics} \textbf{2010}, \emph{43}\relax
\mciteBstWouldAddEndPuncttrue
\mciteSetBstMidEndSepPunct{\mcitedefaultmidpunct}
{\mcitedefaultendpunct}{\mcitedefaultseppunct}\relax
\EndOfBibitem
\bibitem[{K Varela, L R Hargreaves, K Ralphs, M A Khakoo, C Winstead, V McKoy, T N Rescigno and A E Orel}(2015)]{Varela2015}
{K Varela, L R Hargreaves, K Ralphs, M A Khakoo, C Winstead, V McKoy, T N Rescigno and A E Orel} {Excitation of the 4 lowest electronic transitions in methanol by low-energy electrons}. \emph{Journal of Physics B: Atomic, Molecular and Optical Physics} \textbf{2015}, \emph{48}, 115208\relax
\mciteBstWouldAddEndPuncttrue
\mciteSetBstMidEndSepPunct{\mcitedefaultmidpunct}
{\mcitedefaultendpunct}{\mcitedefaultseppunct}\relax
\EndOfBibitem
\bibitem[{K.L. Nixon, W.A.D. Pires, R.F.C. Neves, H.V. Duque, D.B. Jones, M.J. Brunger, M.C.A. Lopes}(2016)]{Nixon2016}
{K.L. Nixon, W.A.D. Pires, R.F.C. Neves, H.V. Duque, D.B. Jones, M.J. Brunger, M.C.A. Lopes} {Electron impact ionisation and fragmentation of methanol and ethanol}. \emph{International Journal of Mass Spectrometry} \textbf{2016}, \emph{404}, 48–59\relax
\mciteBstWouldAddEndPuncttrue
\mciteSetBstMidEndSepPunct{\mcitedefaultmidpunct}
{\mcitedefaultendpunct}{\mcitedefaultseppunct}\relax
\EndOfBibitem
\bibitem[{M Vinodkumar, C Limbachiya, K N Joshipura, B Vaishnav and S Gangopadhyay}(2008)]{Vinodkumar2008}
{M Vinodkumar, C Limbachiya, K N Joshipura, B Vaishnav and S Gangopadhyay} {Computation of total electron scattering cross sections for molecules of astrophysical relevance}. \emph{Journal of Physics: Conference Series} \textbf{2008}, \emph{115}, 012013\relax
\mciteBstWouldAddEndPuncttrue
\mciteSetBstMidEndSepPunct{\mcitedefaultmidpunct}
{\mcitedefaultendpunct}{\mcitedefaultseppunct}\relax
\EndOfBibitem
\bibitem[{M. D. Burrows, S. R. Ryan, W. E. Lamb, Jr., L. C. McIntyre, Jr.}(1979)]{Burrows1979}
{M. D. Burrows, S. R. Ryan, W. E. Lamb, Jr., L. C. McIntyre, Jr.} {Studies of H$^+$, H$_2^+$, and H$_3^+$ dissociative ionization fragments from methane, ethane, methanol, ethanol, and some deuterated methanols using electron‐impact excitation and a time‐of‐flight method incorporating mass analysis}. \emph{The journal of Chemical Physics} \textbf{1979}, \emph{71}, 4931–4940\relax
\mciteBstWouldAddEndPuncttrue
\mciteSetBstMidEndSepPunct{\mcitedefaultmidpunct}
{\mcitedefaultendpunct}{\mcitedefaultseppunct}\relax
\EndOfBibitem
\bibitem[{Sankar De, Jyoti Rajput, A Roy, P N Ghosh, C P Safvan}(2006)]{Sankar2006}
{Sankar De, Jyoti Rajput, A Roy, P N Ghosh, C P Safvan} {Formation of H$_3^+$ due to intramolecular bond rearrangement in doubly charged methanol}. \emph{Physical Review Letters} \textbf{2006}, \emph{97}, 213201\relax
\mciteBstWouldAddEndPuncttrue
\mciteSetBstMidEndSepPunct{\mcitedefaultmidpunct}
{\mcitedefaultendpunct}{\mcitedefaultseppunct}\relax
\EndOfBibitem
\bibitem[{Nagitha Ekanayake, Muath Nairat, Balram Kaderiya, Peyman Feizollah, Bethany Jochim, Travis Severt, Ben Berry, Kanaka Raju Pandiri, Kevin D. Carnes, Shashank Pathak, Daniel Rolles, Artem Rudenko, Itzik Ben-Itzhak, Christopher A. Mancuso, B. Scott Fales, James E. Jackson, Benjamin G. Levine \& Marcos Dantus}(2017)]{Dantus2017}
{Nagitha Ekanayake, Muath Nairat, Balram Kaderiya, Peyman Feizollah, Bethany Jochim, Travis Severt, Ben Berry, Kanaka Raju Pandiri, Kevin D. Carnes, Shashank Pathak, Daniel Rolles, Artem Rudenko, Itzik Ben-Itzhak, Christopher A. Mancuso, B. Scott Fales, James E. Jackson, Benjamin G. Levine \& Marcos Dantus} {Mechanisms and time-resolved dynamics for trihydrogen cation (H$_3^+$) formation from organic molecules in strong laser fields}. \emph{Scientific Reports} \textbf{2017}, \emph{7}, 4703\relax
\mciteBstWouldAddEndPuncttrue
\mciteSetBstMidEndSepPunct{\mcitedefaultmidpunct}
{\mcitedefaultendpunct}{\mcitedefaultseppunct}\relax
\EndOfBibitem
\bibitem[{Nagitha Ekanayake, Travis Severt, Muath Nairat, Nicholas P. Weingartz, Benjamin M. Farris, Balram Kaderiya, Peyman Feizollah, Bethany Jochim, Farzaneh Ziaee, Kurtis Borne, Kanaka Raju P., Kevin D. Carnes, Daniel Rolles, Artem Rudenko, Benjamin G. Levine, James E. Jackson, Itzik Ben-Itzhak \& Marcos Dantus}(2018)]{Dantus2018}
{Nagitha Ekanayake, Travis Severt, Muath Nairat, Nicholas P. Weingartz, Benjamin M. Farris, Balram Kaderiya, Peyman Feizollah, Bethany Jochim, Farzaneh Ziaee, Kurtis Borne, Kanaka Raju P., Kevin D. Carnes, Daniel Rolles, Artem Rudenko, Benjamin G. Levine, James E. Jackson, Itzik Ben-Itzhak \& Marcos Dantus} {H$_2$ roaming chemistry and the formation of H$_3^+$ from organic molecules in strong laser fields}. \emph{Nature Communications} \textbf{2018}, \emph{9}, 5186\relax
\mciteBstWouldAddEndPuncttrue
\mciteSetBstMidEndSepPunct{\mcitedefaultmidpunct}
{\mcitedefaultendpunct}{\mcitedefaultseppunct}\relax
\EndOfBibitem
\bibitem[{Krishnendu Gope, Ester Livshits, Dror M. Bittner, Roi Baer, and Daniel Strasser}(2020)]{Strasser2020}
{Krishnendu Gope, Ester Livshits, Dror M. Bittner, Roi Baer, and Daniel Strasser} Absence of Triplets in Single-Photon Double Ionization of Methanol. \emph{The Journal of Physical Chemistry Letters} \textbf{2020}, \emph{11}, 8108--8113\relax
\mciteBstWouldAddEndPuncttrue
\mciteSetBstMidEndSepPunct{\mcitedefaultmidpunct}
{\mcitedefaultendpunct}{\mcitedefaultseppunct}\relax
\EndOfBibitem
\bibitem[{}()]{NIST}
{} {P.J. Linstrom and W.G. Mallard, Eds., NIST Chemistry WebBook, NIST Standard Reference Database Number 69, National Institute of Standards and Technology, Gaithersburg MD, 20899, https://doi.org/10.18434/T4D303, (retrieved May 17, 2024).} \relax
\mciteBstWouldAddEndPunctfalse
\mciteSetBstMidEndSepPunct{\mcitedefaultmidpunct}
{}{\mcitedefaultseppunct}\relax
\EndOfBibitem
\bibitem[Douglas and Price(2009)Douglas, and Price]{douglas2009studies}
Douglas,~K.~M.; Price,~S.~D. Studies of the fragmentation of the monocation and dication of methanol. \emph{The Journal of chemical physics} \textbf{2009}, \emph{131}\relax
\mciteBstWouldAddEndPuncttrue
\mciteSetBstMidEndSepPunct{\mcitedefaultmidpunct}
{\mcitedefaultendpunct}{\mcitedefaultseppunct}\relax
\EndOfBibitem
\bibitem[{J.H.D. Eland, B.J. Treves-Brown}(1992)]{Brown1992}
{J.H.D. Eland, B.J. Treves-Brown} {The fragmentation of doubly charged methanol}. \emph{International Journal of Mass Spectrometry and Ion Processes} \textbf{1992}, \emph{113}, 167--176\relax
\mciteBstWouldAddEndPuncttrue
\mciteSetBstMidEndSepPunct{\mcitedefaultmidpunct}
{\mcitedefaultendpunct}{\mcitedefaultseppunct}\relax
\EndOfBibitem
\bibitem[{}()]{AIE}
{} {Lide, David R., ed. (2003–2004). "Ionization Potentials of Atoms and Atomic Ions – Section 10. Atomic, Molecular, and Optical Physics". CRC Handbook of Chemistry and Physics (84th edition) (PDF). CRC Press. pp. 10-178. Retrieved 3 December 2020.} \relax
\mciteBstWouldAddEndPunctfalse
\mciteSetBstMidEndSepPunct{\mcitedefaultmidpunct}
{}{\mcitedefaultseppunct}\relax
\EndOfBibitem
\bibitem[{Gordon R. Burton, Wing Fat Chan, Glyn Cooper, C.E. Biron}(1992)]{Burton1992}
{Gordon R. Burton, Wing Fat Chan, Glyn Cooper, C.E. Biron} {Absolute oscillator strengths for photoabsorption (6–360 eV) and ionic photofragmentation (10–80 eV) of methanol}. \emph{Chemical Physics} \textbf{1992}, \emph{163}\relax
\mciteBstWouldAddEndPuncttrue
\mciteSetBstMidEndSepPunct{\mcitedefaultmidpunct}
{\mcitedefaultendpunct}{\mcitedefaultseppunct}\relax
\EndOfBibitem
\bibitem[Guo \latin{et~al.}(1998)Guo, Li, Nibarger, and Gibson]{guo1998single}
Guo,~C.; Li,~M.; Nibarger,~J.~P.; Gibson,~G.~N. Single and double ionization of diatomic molecules in strong laser fields. \emph{Physical Review A} \textbf{1998}, \emph{58}, R4271\relax
\mciteBstWouldAddEndPuncttrue
\mciteSetBstMidEndSepPunct{\mcitedefaultmidpunct}
{\mcitedefaultendpunct}{\mcitedefaultseppunct}\relax
\EndOfBibitem
\bibitem[Rajgara \latin{et~al.}(2003)Rajgara, Krishnamurthy, and Mathur]{rajgara2003electron}
Rajgara,~F.; Krishnamurthy,~M.; Mathur,~D. Electron rescattering and the dissociative ionization of alcohols in intense laser light. \emph{The Journal of chemical physics} \textbf{2003}, \emph{119}, 12224--12230\relax
\mciteBstWouldAddEndPuncttrue
\mciteSetBstMidEndSepPunct{\mcitedefaultmidpunct}
{\mcitedefaultendpunct}{\mcitedefaultseppunct}\relax
\EndOfBibitem
\end{mcitethebibliography}

\end{document}


\maketitle

\begin{figure}
    \centering
    \includegraphics[width=\textwidth]{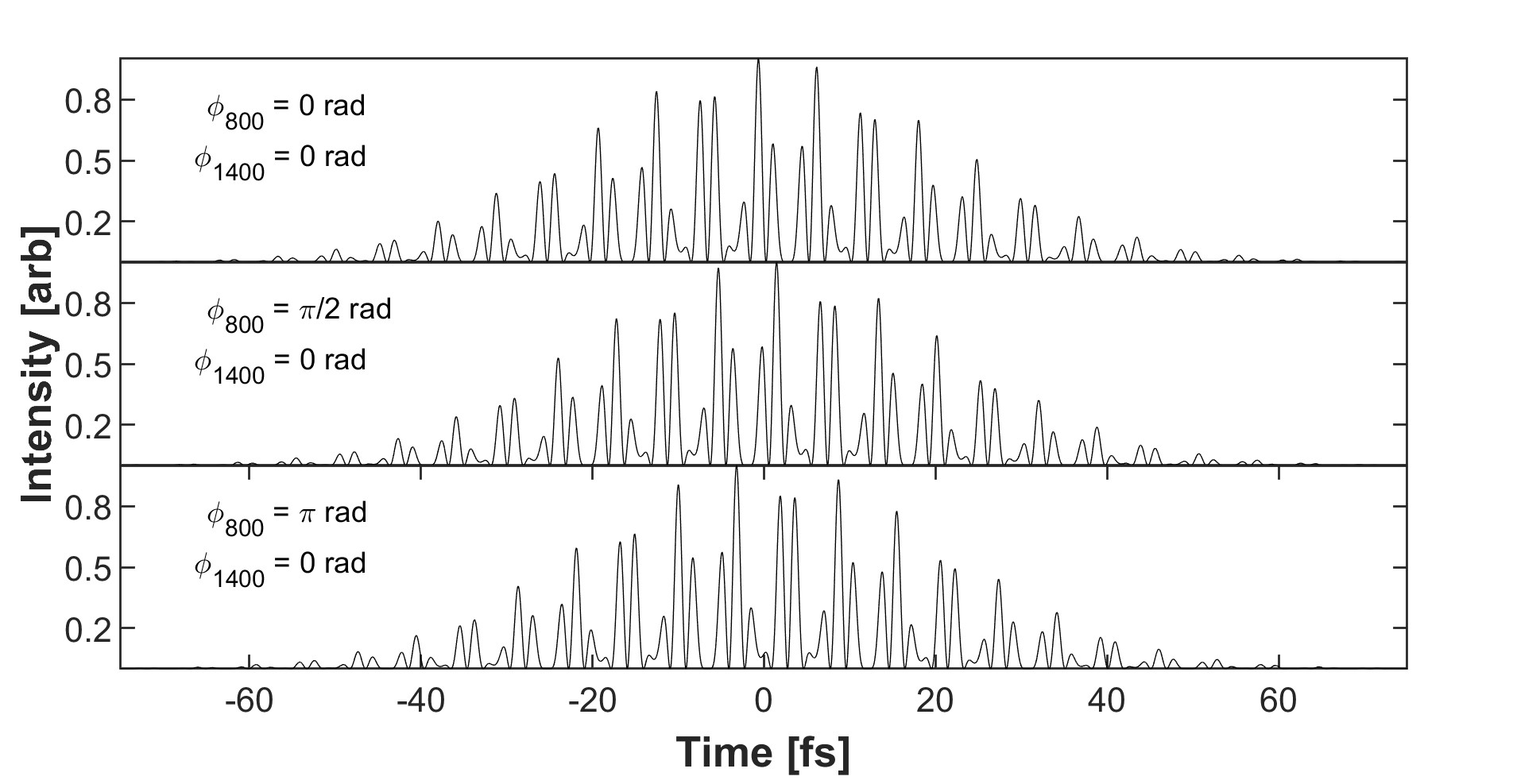}
    \caption*{\textbf{Supplementary Figure 1:} Three NC intensity profiles for different phase delays between the 800 and 1400 nm fields. Note how separated spikes appear in all three cases.}
    \label{figS1}
\end{figure}

\begin{figure}
    \centering
    \includegraphics[width=\textwidth]{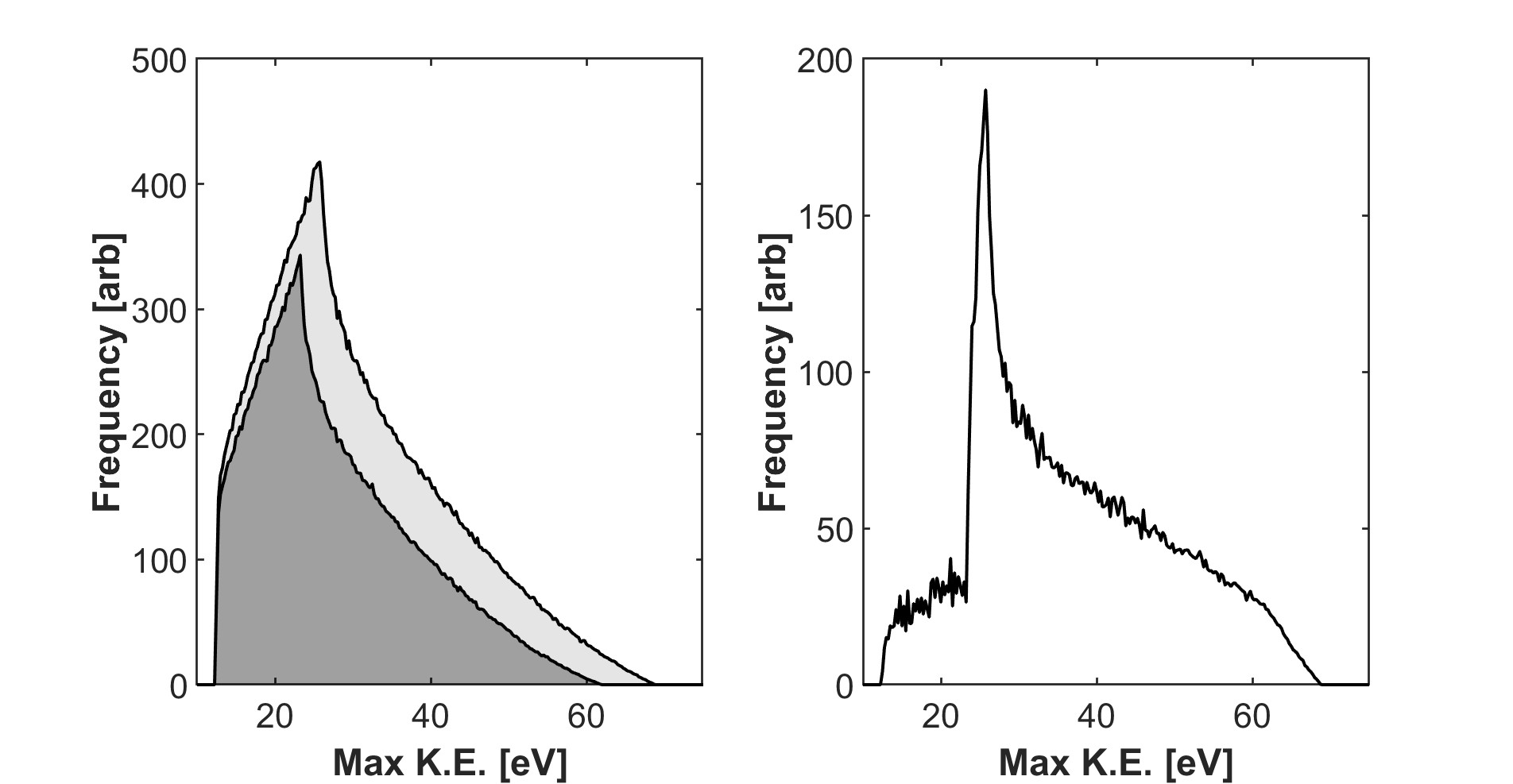}
    \caption*{\textbf{Supplementary Figure 2:} Intensity-difference spectra (IDS) methodology applied to Figure 1(b) of the main manuscript. The left panel shows the calculated electron kinetic energy distribution upon rescattering for the NC fields involving the combination of 800 nm (2.5$\times$10$^{13}$ W cm$^{-2}$) and 1400 nm (7.5$\times$10$^{13}$ W cm$^{-2}$) fields (lighter shaded curve) and 80\% the intensity of both wavelengths (darker shaded curve). The right panel shows the difference between the electron kinetic energy distributions in the left panel, showing that the lower kinetic energy components are significantly reduced.}
    \label{figS2}
\end{figure}

\begin{figure}
    \centering
    \includegraphics[width=\textwidth]{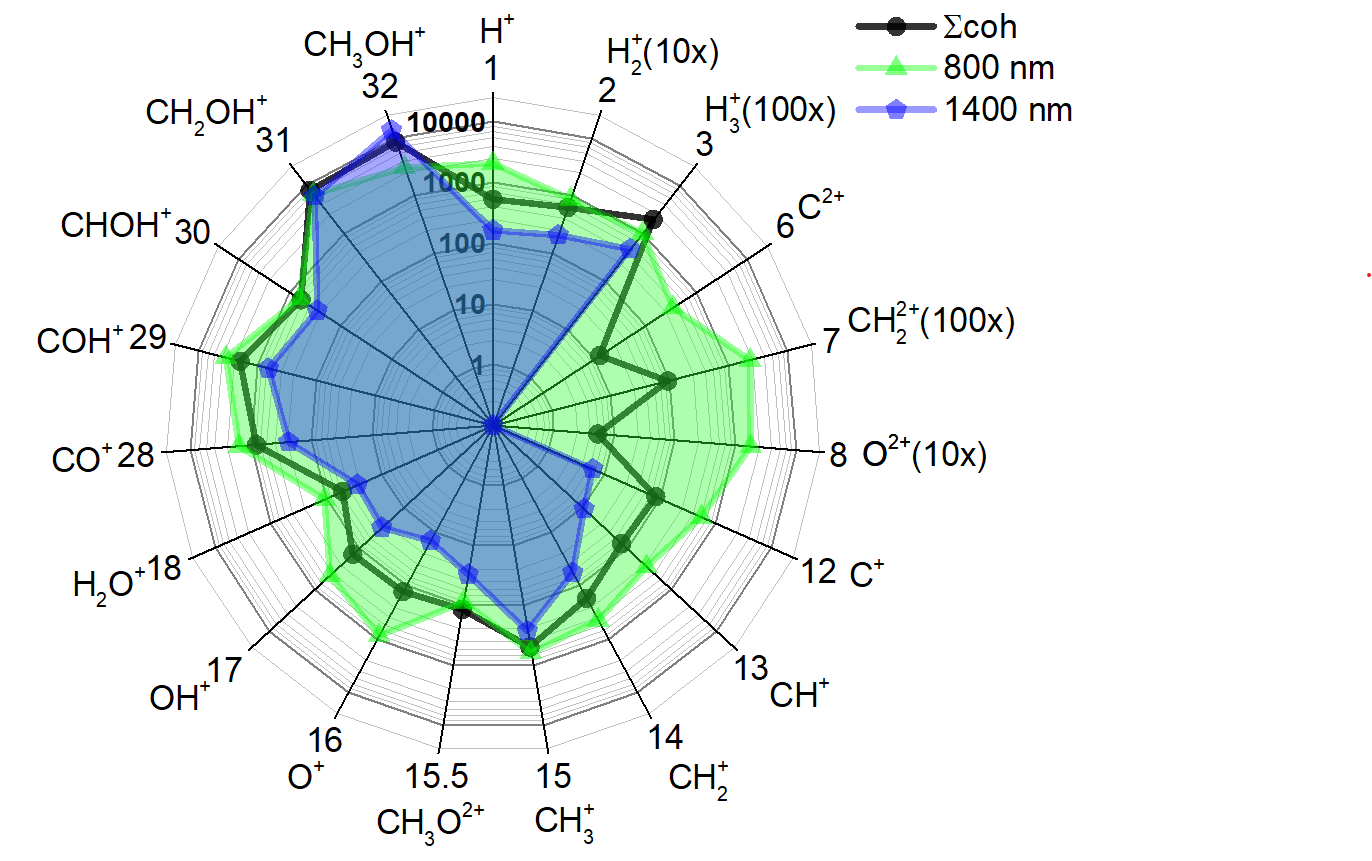}
    \caption*{\textbf{Supplementary Figure 3:} Comparison of enhancement using NC field and individual beams with the same total ion yield. NC field ($\Sigma$coh) - 1400 nm at 3.9 $\times$ 10$^{14}$ W cm$^{-2}$ and 800 nm at 9.1 $\times$ 10$^{13}$ W cm$^{-2}$, 800 nm - 2.6 $\times$ 10$^{14}$ W cm$^{-2}$, 1400 nm - 6.7 $\times$ 10$^{14}$ W cm$^{-2}$.}
    \label{figS3}
\end{figure}